\documentclass[]{interact}
\usepackage{graphicx}
\usepackage{epstopdf}
\usepackage[caption=false]{subfig}
\usepackage{float}
\usepackage{tikz}
\usetikzlibrary{calc,shapes,backgrounds, arrows, positioning, fit}

\usepackage[natbibapa,nodoi]{apacite}
\theoremstyle{plain}

\theoremstyle{definition}

\theoremstyle{remark}

\begin{document}

\articletype{ORIGINAL ARTICLE}

\title{Analyzing Ecological Momentary Assessment Data with State-Space Models: Considerations and Recommendations}

\author{
Lindley R. Slipetz \\
\affil{University of Virginia, Charlottesville, USA}
Jeremy W. Eberle\\
\affil{University of Virginia, Charlottesville, USA}
Cheri A. Levinson\\
\affil{University of Louisville, Louisville, USA}
Ami Falk\\
\affil{University of Virginia, Charlottesville, USA}
Claire E. Cusack\\
\affil{University of Louisville, Louisville, USA}
Teague R. Henry\\
\affil{University of Virginia, Charlottesville, USA}
}

\maketitle

\begin{abstract}
Ecological momentary assessment (EMA) data have a broad base of application in the study of time trends and relations. In EMA studies, there are a number of design considerations which influence the analysis of the data. One general modeling framework is particularly well-suited for these analyses: state-space modeling. Here, we present the state-space modeling framework with recommendations for the considerations that go into modeling EMA data. These recommendations can account for the issues that come up in EMA data analysis such as idiographic versus nomothetic modeling, missing data, and stationary versus non-stationary data. In addition, we suggest R packages in order to implement these recommendations in practice. Overall, well-designed EMA studies offer opportunities for researchers to handle the momentary minutiae in their assessment of psychological phenomena. 

\end{abstract}


\maketitle
\newpage
Ecological momentary assessment (EMA) is a type of timeseries collection technique in which measures are typically gathered multiple times a day over several weeks in the participant's natural environment. Longitudinal data has a variety of subtypes, including panel data and intensive longitudinal data. Panel data traditionally has a greater sample size and smaller number of timepoints than intensive longitudinal data, and EMA can most appropriately be described as the latter. Due to the complexities of EMA data, that is, dependencies between measured values, missingness, and other design/modeling decision points, correct modeling in this medium is a challenge. In this paper, we consider these researcher decision points and present design and modeling solutions to the obstacles faced by applied researchers studying EMA data. By utilizing the state-space model, a general modeling framework, we show how to quantitatively account for each of these choices, for example, what type of model a researcher should use depending on the variability of when data was collected. Overall, the aim is to guide researchers in the use of EMA and state-space modeling. 

As opposed to EMA studies, in-lab experiments or assessments offer a snapshot of the phenomenon under study and do not take into account changes in the days and weeks prior to assessment \citep{wright_relationship_2021}. Such assessments also rely on the participant's memory and/or their ability to make generalizations about previous experiences \citep{mote_ecological_2020}. For example, in a clinical interview, responding to the item "In the past two weeks, were you fatigued when you felt depressed?" requires the participant to remember the instances of depression and fatigue from the past two weeks and judge if there was a general relationship between the two over this period of time. In addition, in-lab assessments have the further problem of contextual influences in which trust, cultural biases, and others play a role in the participant's comfort in disclosing \citep{wright_relationship_2021}. EMA data offers an alternative to the in-lab assessment approach which can overcome these hurdles.  

 At its core, EMA is a data collection protocol for assessing human thoughts, feelings, behaviors and even physiology in participants' lived experience. Participant responses are collected in real time/near real time as the participants go about their daily lives, typically via an application on the participant's smartphone. This design minimizes the need for recall and allows for precise timing information to be collected. Theoretically, EMA has less bias than in-lab assessments as, for instance, it reduces the need to remember or generalize over multiple experiences \citep{mote_ecological_2020}. For example, instead of responding to "In the past two weeks, were you fatigued when you felt depressed?," a participant in an EMA study would be asked about their fatigue and depression multiple times a day for several weeks.
 
 EMA collection is either device-prompted (e.g., the data collection device prompts the user to answer questions at particular times of day) or participant-initiated (e.g., an event occurs and the participant logs it) \citep{srivastava_momentary_2021}. For instance, participants might be instructed to record instances of specific behavior (e.g., when they smoke a cigarette or drink alcohol), and might be prompted to provide more information (e.g., current affective state), or participants might be prompted (that is, “pinged”) to respond multiple times over the course of a day, and this might include different questionnaires for different times of day (e.g., a brief affective state questionnaire during the middle of the day with a longer battery of items administered at the nightly ping). Another kind of participant-initiated schedule is location-initiated, in which participants might be asked to respond if they go to specific locations (e.g., in a study of alcohol use, bars). In addition to participant-initiated and scheduled data collection, if ambulatory psychophysiology is being collected, then there may be continuous physiological monitoring throughout the day and night.

Once collected, EMA data provide a rich substrate for analyses of time trends and relations. For example, in a study of alcohol use, a researcher might be interested in the relation between alcohol use and the time of day or day of the week. Depending on the approach taken by the researcher the results from an analysis of EMA data can focus inference on substantially different aspects of the data. The gamut of possible uses and analyses of EMA data can be seen via the following examples:

\citet{ditzen_positive_2008} studied the relationship between intimacy and daily salivary cortisol levels in order to assess protective health factors in couples. Their EMA study used self-reports of time spent on intimacy, chronic problems at work, and affect quality, and saliva samples for cortisol level estimation were collected every three hours for one week. Notably, the EMA protocol combined the behavioral/psychological measurement via smartphone that is common to modern EMA studies with saliva samples that the participants provided. In particular, saliva samples were provided at the same sampling rate as the behavioral measures, making it possible to analyze the relation between cortisol and behavior at the same timescales. While the physical collection of saliva samples would increase participant burden, this study demonstrates that it is possible to use an EMA-style protocol while obtaining more difficult-to-collect types of data. 

\citet{mackerron_happiness_2013} sought to assess the link between well-being and physical environment. In this EMA study, participants were pinged randomly during a time interval and frequency of their choosing. They were asked their level of happiness on a continuous sliding scale, whom they were with, where they were, and what they were doing. While filling out the survey, the precise location of the individual was found via GPS. The GPS data was then associated with three indicators: habitat/land type (e.g., green space), weather conditions, and daylight status. This study is notable for its scale and use of GPS location data.  While EMA data collection is often associated with small sample sizes, this study demonstrates that EMA can be done on a large scale: this study had 1,138,481 data points from 21,947 UK participants. It also demonstrates the use of passive data collection: by collecting GPS coordinates, the authors were able to study the association between affect and location without relying on potentially unreliable retrospective self-report of location.

An EMA protocol can also be combined with experimental methods to examine, for example, the impact of an intervention on the moment-to-moment dynamics of behaviors. In a recent clinical trial by \citet{levinson_personalizing_2023}, EMA data collection was used to identify person-specific central symptoms of eating disorders, which were then used to inform a personalized multi-week treatment plan. Here, EMA data was collected before the intervention, while the efficacy of the intervention was evaluated with more traditional single measurement occasion questionnaires. In general, there is no barrier beyond participant burden to collecting EMA data before, during, and after an intervention to track how behavioral dynamics change in what effectively amounts to real time. EMA data collection can also form part of an intervention itself. These applications of an EMA protocol go by the term ecological momentary interventions  \citep[EMI;][]{mcdevitt_murphy_use_2018} and have shown promise in the treatment of a number of psychological disorders/behaviors, most notably depression \citep{schueller_ecological_2017}. For example, in \citet{levinson_personalizing_2023}, if a participant was found to have fear of rejection as a central cognition symptom, the intervention would be social exposure, followed by further EMA collection and interventions on other central symptoms.

In summary, when contrasted with in-laboratory experiments or panel longitudinal studies, there are two key advantages to EMA. First, there is improved ecological validity; that is, with EMA data, the behaviors and responses are being recorded in the participant's naturalistic environment. Second, EMA data provides fine-grained timing information which allows researchers to analyze the within-day dynamics of behaviors or psychological states. This cannot be done with traditional panel longitudinal designs.

However, even with the promise of EMA data, there are a number of design and analysis considerations that impact how researchers can analyze this data. In this article, we first review a general modeling framework, state-space modeling, that is particularly well suited for the analysis of EMA data. State-space modeling is a type of dynamic latent variable model, in which the unobserved state variables track the trajectory of the phenomenon under study. Next, we describe a number of considerations regarding EMA studies and data, their consequences both analytical and inferential, and related approaches that one can use to account for these issues.
 
\section{State-Space Models for EMA Data}

Used in the analysis of time series data, the state-space modeling framework is a general framework that represents the observed time-varying variables as measurements of a set of underlying, unobserved \textit{state} variables \citep{kalman_new_1960}. State-space models integrate classical psychometric measurement (i.e., confirmatory factor analysis or CFA) with temporal dynamics of the unobserved latent states. For example, in a CFA model of depression, the latent variable depression is measured by a single selection of items from a depression questionnaire. Expanding this over time, a EMA study can collect repeated within-person measurements of that same selection of items from a depression questionnaire, where depression would be a latent state measured over time by the questionnaire responses. A state-space modeling framework would allow the researcher to both account for the measurement of depression, and assess how depression is related to itself over time. 

As an example of state-space modeling applied to psychological data, \citet{mckee_emotion_2020} presented a feedback-control model of emotional homeostasis in order to arrive at an index of emotion regulation for a time series of 11 affect items over 18 timepoints for a high-risk adolescent (i.e., ages 13-14) sample. In this continuous time model, the state was affect, which was measured by the observed affect measures with an autoregressive regulation parameter as the index of emotion regulation (which was defined as the percent recovery to equilibrium over each time period) and a standard deviation of exogenous disturbances (which was defined as the unmeasured influences on the individuals). A cross-sectional between-person analysis of total counts of nicotine, alcohol, and cannabis use was performed with the regulation parameter, standard deviation of exogenous disturbances, and means/standard deviations of sum scores for the measures as predictors. They found significant relationships between affective state and substance use, perhaps indicating bidirectional feedback cycles in which, for example, higher negative affect was associated with a higher tendency to smoke cigarettes and lesser standard deviations of mood were associated with a higher tendency to smoke cigarettes. What this study demonstrates is the ability of state-space models to track affective regulation over time, specifically, or regulation of psychological phenomena, generally, over time.

Perhaps the simplest, most used, and most studied state-space model is the discrete time multivariate normal state-space model for a single participant's time series:
\begin{align}
    \mathbf{x}_{t+1} &= \mathbf{A}\mathbf{x}_t + \mathbf{G}\mathbf{u}_t + \boldsymbol{\varepsilon}_t \\
    \mathbf{y}_t &= \mathbf{H} \mathbf{x}_t + \boldsymbol{\nu}_t \\
    \boldsymbol{\varepsilon}_t & \sim N(0, \boldsymbol{\Sigma})\\
    \boldsymbol{\nu}_t & \sim N(0, \boldsymbol{\Theta})
\end{align}
where $x_t$ is a vector of values for the unobserved states at time $t$ (e.g., depression), $u_t$ is a vector of observed exogenous disruptions (e.g., an external event like a death in the family or an intervention like social exposure above), $y_t$ is a vector of observed measurements at time $t$ (e.g., items on the Beck Depression Inventory), $\mathbf{A}$ is a square matrix of coefficients that describe how the state values are related across time (e.g., autoregressive relationships between depression at time $t$ and at $t+1$), $\mathbf{G}$ is a matrix that maps the effects of the exogenous disturbances (e.g., the effect of a scheduled intervention) onto the states, and $\mathbf{H}$ is a matrix of coefficients that describe how each observed measure maps onto the underlying state values (i.e., like factor loadings in CFA). $\boldsymbol{\varepsilon}_t$ and $\boldsymbol{\nu}_t$ are the state innovations (i.e., the over-time variability associated with the states) and measurement error variables respectively, and their distribution is multivariate normal of the appropriate dimension with expected values of 0 and covariance matrices $\boldsymbol{\Sigma}$, $\boldsymbol{\Theta}$ respectively. 

Observant readers will notice that the model described in Equations 1-4 is similar to both classical vector autoregressive (i.e., VAR) models, as well as to latent variable models commonly used throughout psychological and behavioral sciences. The state-space framework is very general and many standard models can be described using state-space notation. For example, the above model can be reduced to a VAR model of order 1 (i.e., the model is defined over only one lag) by considering $x_t$ to be observed, and ignoring $y$ completely. For the similarities to latent variable models, the state variables $x$ are latent variables, $y$ are manifest indicators, the $\mathbf{H}$ matrix can be considered a matrix of factor loadings (often denoted as $\boldsymbol{\Lambda}$ in the structural equation modeling literature), $\mathbf{A}$ is analogous to the matrix of inter-factor regression coefficients (often $\mathbf{B}$), and  $\boldsymbol{\varepsilon}_t$, $\boldsymbol{\nu}_t$ are the error distributions of the latent variables and indicators respectively. Most models representing the temporal dynamics of variables can be expressed as a state-space model. With the exception of measurement, which entails an assumption of an underlying latent state, none of the issues or considerations we describe in this article are dependent on the presence of latent states, and are applicable to the broader class of multivariate time-series models such as VAR models. Nevertheless, we focus on the state-space framework because of its generality and applicability to a wide range of EMA data. 

It must also be noted that the model described by Eqs 1-4 has a number of strong assumptions associated with it, but these assumptions can be relaxed/changed and are not intrinsic to the state-space modeling framework. For example, the model operates over discrete time, in that each time interval from $t$ to $t+1$ is considered to be equivalent. The important part about the lags between the measurements in discrete models is not the length of each lag, but that each is equivalent. The model is also time invariant, in that the various model parameters do not vary over the study period. In addition, the model describes the dynamics for a single multivariate timeseries, rather than describing the aggregate dynamics for multiple participants' worth of timeseries data. Finally, the model assumes multivariate normality of both states and observations, which brings with it the assumption that all temporal dynamics between states can be described linearly. Again, none of these assumptions are intrinsic to the state-space modeling framework. Throughout the remainder of this paper, we will show how the above model can be modified (i.e., by relaxing/changing assumptions) to account for various features of EMA data.

The remainder of this article is structured as follows: We describe a number of considerations that we believe scientists planning EMA studies or analyzing EMA data should be familiar with and account for in their design/analysis. In each case, we will discuss what the consideration is, why it matters to the analysis of EMA data, and, if possible, how the model described in Equations 1-4 can be modified to account for the consideration. While we will not describe the specifics of estimating these models, we will provide references to appropriate R packages, as well as references for further reading on various state-space modeling approaches. Note that we explicitly and only cite publicly available R packages, both to reduce the scope of review and to encourage the use of open-source and freely available scientific software.

\section{Considerations and Recommendations}

\subsection{Idiographic vs. Nomothetic Modeling}

EMA collects rich within-participant data, often resulting in a large number of observations per participant. The first aspect of modeling researchers should consider when weighing analysis strategies is whether they think the phenomena under study will differ wildly across participants, be the same across participants, or lie somewhere between those two extremes. This is the consideration of idiographic versus nomothetic modeling \citep{silverstein_aristotelian_1988}. 

Consider an EMA study in which the dynamics of depression, a highly heterogeneous disorder \citep{fried_depression_2015}, are being modelled. A purely idiographic approach implies that each individual has a unique set of symptom dynamics; a partially idiographic approach implies that some pattern of dynamics is common across the sample, but there is inter-individual variability; while a fully nomothetic approach implies that all individuals have the exact same dynamics underlying their depression symptoms. Clearly, the choice of an idiographic vs. nomothetic approach is strongly related to the theory behind the phenomena of study.

\subsubsection{Purely idiographic and nomothetic approaches}

There are dangers to incorrectly applying a nomothetic approach when the phenomenon is fully or partially idiographic or applying an idiographic approach when a nomothetic approach is more appropriate. In the former case, the nomothetic model would likely not represent any one participant well, and in the worst case scenario of a purely idiographic process, not represent the dynamics of any participant \citep{molenaar_manifesto_2004}. On the other hand, in the latter case, applying an idiographic approach when the phenomenon has the same dynamics across all individuals will result in lost power and bias in estimation, as information is not being pooled across individuals \citep{gelman_multilevel_2006}. Because in the nomothetic case it is likely the dynamics will not be appropriately represented, we recommend that researchers err on the side of idiographic analysis, unless there is a compelling theoretical reason to believe that the phenomenon is nomothetic.

The two extremes, when a purely idiographic approach is appropriate versus when a purely nomothetic approach is appropriate, are simple to represent in the state-space modeling framework. For the purely idiographic approach, each individual's timeseries would be modeled separately, resulting in unique estimates of the various model matrices ($\mathbf{A},\mathbf{G},\mathbf{H},\mathbf{\Sigma}, \mathbf{\Theta}$) for each individual. Similarly, in the purely nomothetic case, individual timeseries can be concatenated (with breaks between participants' data being represented) and a single model fit, estimating a single set of model matrices that, in theory, represent the dynamics for every participant in the sample. Both of these cases can be estimated using common state-space modeling packages such as \texttt{dlm} \citep{petris_r_2010} or \texttt{dynr} \citep{ou_r_2019} without much difficulty. Both packages have the ability to fit discrete time state-space models using the Kalman filter (see Missing Data section).

\subsubsection{Partial idiographic approaches}

More difficult is the partial idiographic approach. Here, a common set of dynamics hold across individuals, but individuals are allowed to have their own variations on the group dynamics. One issue here is that the notion of individual variation "around" group dynamics is ill-defined. For example, individuals could have the same set of structural parameters (i.e., non-zero entries in the $\mathbf{A}$ matrix), but those non-zero parameters can take on different values. Alternatively, there could be a common set of relations for the entire sample (i.e. the cross-lagged relation between depression and anxiety states), with an additional set of relations per individual (i.e. in some individuals, there could be a relation between anxiety and impulsivity, but not for all analyzed individuals).

There are several models/software packages that allow for partial idiographic modeling of multivariate timeseries data. The \texttt{gimme} \citep{lane_2022} package implements the \textit{Group Iterative Multiple Model Estimation} algorithm \citep{gates_group_2012}, which fits structural VAR models iteratively to multiple time series in order to estimate group paths (i.e., dynamics in the $\mathbf{A}$ matrix that are common across the whole sample) and individual paths. A stepwise model search procedure results in sparse dynamics being estimated for both group and individual paths.  Researchers should use the \texttt{gimme} package when estimating partially idiographic models with latent variables (i.e., traditional state-space) or when concerned with automatic detection and confirmatory testing of subgroups \citep{henry_comparing_2018, lane_uncovering_2018}, though it also can estimate observed variable-only models.  

Alternatively, the \texttt{multivar} package \citep{fisher_2022} provides an implementation of a multi-participant regularized VAR model for observed variables. This approach allows researchers to estimate group and individual effects combined with the regularization's ability to estimate sparse (i.e., some entries in the $\mathbf{A}$ matrix are set to 0) dynamics. Researchers should use the \texttt{multivar} package when fitting observed variable multi-participant models with a large number of parameters (as regularization will shrink some to zero, making the model more computationally simple and increasing interpretability). Both GIMME and \texttt{multivar} estimate sparse dynamics, though GIMME estimates sparse dynamics using a model building approach, which leads to group-level effects, and effects that are specific to the individual. 

The \texttt{mlVAR} \citep{epskamp_2021} package implements a multi-level or mixed effects VAR, where random effects are used to model inter-individual variations in observed variable dynamics. Multi-level VAR does not result in sparse dynamics, but does result in a group-level effect and individual-level effects for all parameters (that are specified with a random effect). Researchers should use \texttt{mlVAR} when they have fewer variables and wish to estimate a multiple participant observed variable model. We will note that with the exception of \texttt{gimme}, there is no R package that allows for the partially ideographic models that contain latent states to be estimated. Partial idiographic models for multivariate timeseries are an active field of research, particularly in the health and behavioral sciences where multiple participants worth of timeseries data are regularly collected, and new innovations are being actively developed. For accessible work on idiographic and nomothetic modeling in the case of EMA, we suggest \citet{wright_modeling_2020, fisher_lack_2018, foster_advancing_2018}.

\begin{figure}[H]
\centering
\includegraphics[width=.9\textwidth]{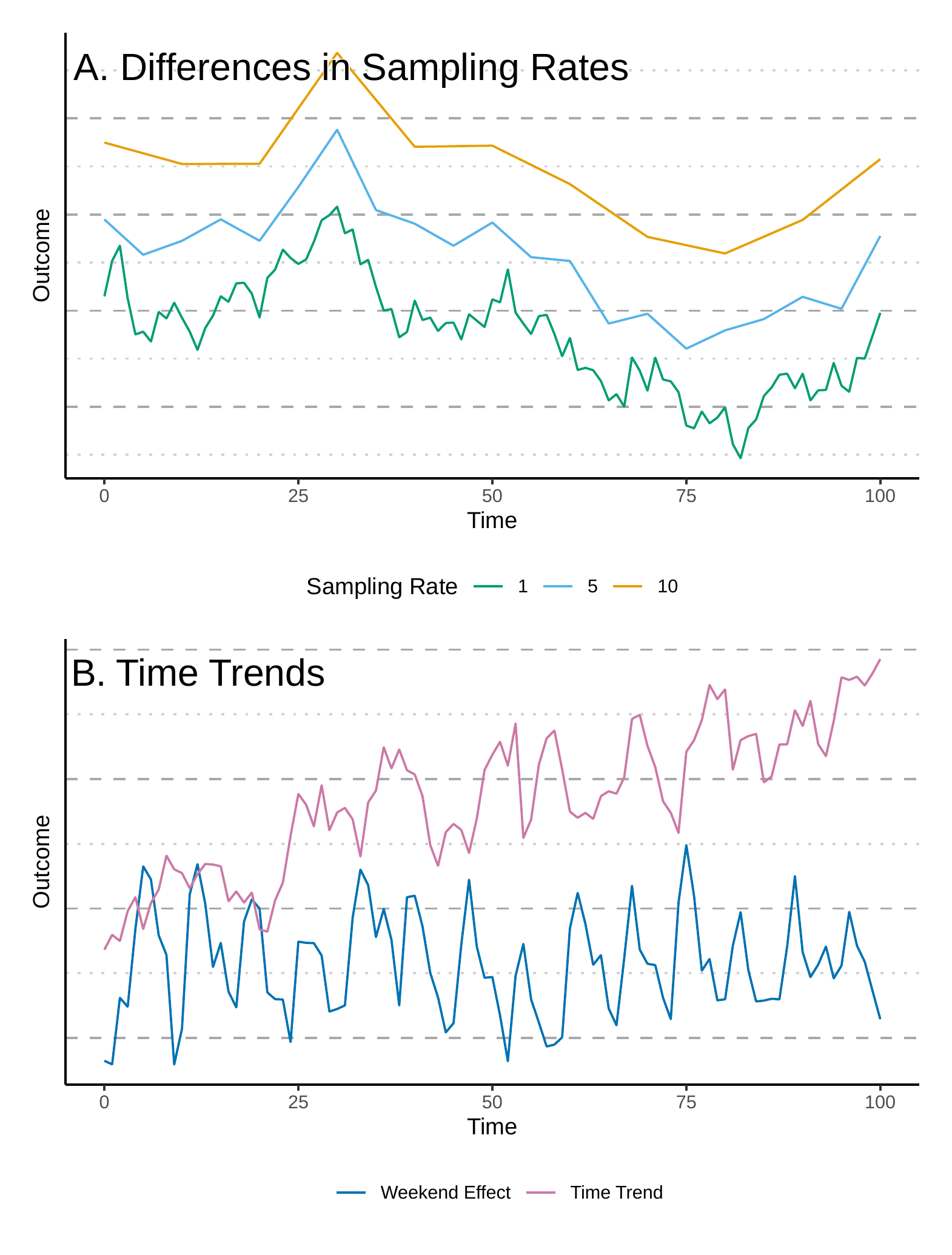}
\caption{Panel A illustrates the relationship between sampling rate and the patterns of the dynamics. The green line is a sampling rate sufficient for capturing the patterns in the dynamics. The blue line reduces the sampling rate to 1 timepoint per every 5 timepoints in the original data. Notice, here, that the blue line fails to capture the smaller fluctuations, yet follows the overall pattern. The orange line is an even further reduction in sampling rate (1 timepoint to every 10 timepoints in the original data). This further reduction results in lesser ability to capture the patterns in the dynamics. Panel B illustrates two types of time trends: a linear increase over time (magenta line) and a weekend effect (blue line). For this graph, the timeseries share the same parameters except for differing in time trends. This demonstrates the strong influence that time trends can have on timeseries.}
\label{fig1}
\end{figure}

\subsection{Data Collection Schedule}

Several study design and analysis considerations may inform the \textit{ping} schedule, or when over the course of the days/weeks/months of the study period are data collected. There are two broad types of ping schedules (which can be combined). First, prescheduled pings, where the prompts for responses are sent to participants or participants are told to record at specific times of the day/week (e.g., each day at noon). Second, event-driven pings, where participants are either prompted to respond when certain conditions are met (such as the participant enters a location as measured by GPS), or, more commonly, the participants are instructed to respond before/during/after a specific behavior or event (e.g., responding after an episode of alcohol use in a study of alcohol use, or responding after a meal in a study of disordered eating). 

\subsubsection{Sampling rate}

There are several concerns at the study design level and at the analysis level to which the ping schedule relates. Most pressingly, data collection must be frequent enough to measure changes in the theoretically relevant process. Figure \ref{fig1} Panel A illustrates how decreasing one's sampling rate can obscure relevant patterns of the dynamics. The green line can be considered data sampled at a rate sufficient to capture all the dynamics. The blue line reduces the sampling rate to 1 timepoint per every 5 timepoints in the original data. Note that with the reduction in the sampling rate, the smaller fluctuations are obscured, while the overall pattern is maintained. Further reductions in the sampling rate (yellow line) increasingly degrades our ability to track fluctuations in the dynamics. When designing an EMA study, researchers should take care in identifying an adequate sampling rate for the phenomena under study. Importantly, this does not mean that every EMA study should have a high sampling rate. In many situations, the process under study changes slowly (e.g., \citet{courtet_circadian_2012} write that there are diurnal mood swings in depression with symptoms being worse in the morning whereas \citet{proudfoot_evidence_2014} found mood to cycle weekly in bipolar disorder). Nyquist's theorem \citep{nyquist_certain_1928} provides a general rule for determining sampling rates: the sampling rate must be more than twice the frequency of the phenomena under study (e.g., if a process changes on the order of 24 hours, researchers should choose to sample on the order of every 12 hours). For event-related phenomena (e.g., substance use), researchers can choose a behavior-driven schedule which will likely result in reasonable accuracy; however such a schedule may need to be augmented with a pre-scheduled setup to capture unrecorded events. For a recent look at the effect of sampling rates on inference in dynamical systems, see \citet{haslbeck_recovering_2022}.

\subsubsection{Discrete versus continuous time}

With respect to analysis, the ping schedule has considerable implications for the performance of a modeling approach. If the pings are approximately equally spaced (e.g., 1 ping per day at approximately the same time of day), then researchers can use the discrete time state-space model described in Eqs 1-4 to model the dynamics. However, if the ping schedule results in highly variable intervals between pings (e.g., with a behavior-driven schedule where the event does not occur at equal intervals or where there is clustering in measurement), then researchers should use a continuous time state-space model. A continuous time version of the model described in Eqs 1-4 changes Eq 1 into a stochastic differential equation:
\begin{equation}
    \dot{\mathbf{x}_t} = \mathbf{A}\mathbf{x}_t + \mathbf{Gu}_t + \boldsymbol{\varepsilon}_t
\end{equation}
where the instantaneous rate of change $\dot{\mathbf{x}_t}$ is a function of the current value of $x$, any disturbance effects of $u$ and a random effect. This model can be considered to be a hidden Markov model in which the states are continuous rather than discrete. Using a discrete time model when a continuous time model is more appropriate can lead to increased bias in effect estimates \citep{de_haan-rietdijk_discrete-_2017}, though some work shows that a discrete approximation can be used to model EMA data, as long as the ping intervals are not highly variable \citep{tuerlinckx_comparison_2001, boker_generalized_2010}. 

There are differences in the interpretation of discrete and continuous time state-space parameter estimates; however, these differences are due to the units of the model, and parameter estimates can be made equivalent between discrete and continuous time cases, at least when linear dynamics are assumed. For example, if a two state discrete state-space model has the state dynamic parameter matrix $  \mathbf{A} = 
	\begin{bmatrix} 
	.5 & .2 \\
	0 & .5 
	\end{bmatrix}$, the equivalent continuous time representation is $  \log{\mathbf{A}} = 
	\begin{bmatrix} 
	-.69 & .4 \\
	0 & -.69 
	\end{bmatrix}$. These parameter matrices have different interpretations, with $\mathbf{A}$ being the relation between the states from time $t$ to time $t+1$, while $\log{\mathbf{A}}$ is the relation between the state values at time $t$ and the first derivative (i.e. the rate of change) at time $t$. Because there is a 1-to-1 relation in that the dynamic parameters of the continuous time model are the exponentiated values of the dynamic parameters of the discrete time model, we can consider the representations equivalent, but warn that this only holds in this fashion for linear models. For a detailed analysis and discussion of the relation between discrete and continuous time model parameters, see \citet{oud_continuous_2007}. 
 
 While there are, of course, differences in interpretation for discrete versus continuous time state-space models, the largest difference is in the methods of estimation, which limits the number of R packages implementing continuous time models. As of writing, the \texttt{dynr} \citep{ou_r_2019}, \texttt{OpenMx} \citep{openmx} and \texttt{ctsem} \citep{driver_2017} packages are the only R packages that implement continuous time state-space models. If researchers suspect there may be regime switching in their model (see section, "Stationarity," below) or they have multiple subject data, the \texttt{dynr} package is suggested. If researchers are interested in mixed effect continuous time models, then we advise using \texttt{ctsem}. Finally, \texttt{OpenMx} is the most general latent variable modeling package of the above listed, and researchers can use it for observed variable models, latent variable models, cross-sectional models, and dynamic models. For accessible work further reviewing continuous time state-space models, we suggest \citet{oud_continuous_2000, oravecz_hierarchical_2011, lodewyckx_hierarchical_2011}.

\subsubsection{Number of bursts}

In addition to considering when and how pings will occur, researchers must consider how many clusters of pings will occur. Many studies take advantage of EMA's benefits by combining it with other study designs arising from a decision point to do a single burst or multiple burst design. For example, a multiple burst protocol is to have an interval of EMA collection before a treatment, then a separate interval of EMA collection after treatment. Alternatively, a single burst design (like the one in \citet{mackerron_happiness_2013}) can be useful when researchers are less concerned with changes over different periods of time. Overall, choice of single or multiple burst is entailed by the research question: like ping schedule, at the study design level, theory-based care should be taken in the decision so as to properly measure the phenomenon under study. For examples of single versus multiple burst designs see \citet{bai_effects_2020, ram_questionable_2017, malmberg_within-student_2016, lay_neuroticism_2017, fuller-tyszkiewicz_state-based_2020}.

Multiple burst designs are simple to analyze, conditional on the proper analysis of the within-burst data, as each burst can be analyzed separately and compared post-analysis. One important consideration is how the process under study might change during the course of treatment or over time. For example, in a study of parents' and adolescents' well-being, \citet{janssen_does_2020} used a multiple burst design of EMA measurement before COVID-19 and during COVID-19, a difference in processes that will be apparent to the contemporary reader. If a researcher thinks that such differences are theoretically possible, then a multiple burst design should be favored. When a researcher thinks that longitudinal differences or pre/post tests are irrelevant, then a single burst is adequate. Overall, the design for number of bursts in data collection design is dependent upon the kinds of changes the researcher expects to see in the phenomena under study. For further reading on single vs. multiple burst designs in EMA studies, see \citet{nesselroade_warp_1991, sliwinski_measurement-burst_2008}.

\subsection{Time Trends and Night Effects}

Until now we have implicitly assumed there are no systematic or cyclic trends with respect to time that effect the process under study. Unfortunately, in the study of thoughts, feelings, and behaviors, there are rarely no systematic trends. Take for example the well known ``weekend effect'' on mood \citep{stone_day--week_2012}, where positive affect is simply greater during the weekend than during weekdays. Another example of a time-related trend is the diurnal rhythm of cortisol, a hormone linked to stress response. Cortisol is the highest immediately following waking and generally declines over the course of the day \citep{desantis_racialethnic_2007}. More to the point, humans have a cyclical rhythm, with periods of activity following periods of sleep, so any EMA study must consider, if not only to dismiss as irrelevant, the effect of sleep on their study outcomes. 

\subsubsection{Time trends}

Panel B in Figure 1 illustrates two types of time trends. The blue line represents a simple linear increase over time, while the magenta line represents an increase during the ``weekend'' of each simulated week. Note that aside from their time trends, the rest of the parameters used to generate these timeseries are the same, which highlights the power that time trends have in shaping the trajectory of the timeseries. 

There are a number of ways that state-space modeling described in Eqs 1-4 can accommodate both time trends. Time can be coded as an exogenous variable and included in $\mathbf{u}_t$. Depending on the manner of coding, this exogenous variable approach can represent a wide range of kinds of time effects. Researchers who hypothesize that there is a linear trend like the one portrayed in Figure 1.B can include the absolute time from the beginning of the study (eg. $t = 1,2,\dots,T$), while researchers who believe theoretically that there are weekend effects (or simply specific periods of time effects) can model them as 0/1 codes. Finally, when researchers believe that there are more continuous fluctuations or within-day effects, these can be modeled by coding time to reflect time of day or time since waking. Care must be taken however, particularly when considering the analysis of multiple studies. \citet{shiffman_ecological_2008} advises that time be coded in a way such that the relative differences in the participant's timeline are accounted for. For example, instead of using the absolute time of day (e.g., time since midnight), to use time since waking. This accounts for different participant schedules. For an accessible overview of many time or event order related issues specific to EMA collection, we recommend a close read of \citet{shiffman_ecological_2008}.

\subsubsection{Night effects}
Here, we use the term "night effects" to capture any differences in the dynamics that might be due to sleep. For example, while affective processes might be fairly stable for a given person during waking hours, it is unlikely that the same stable dynamics occur during sleeping hours (e.g. mood might reset to a baseline during sleep). Night effects are less easy to model than time trends or irregular time intervals, as they are not due to differences in the timing of collected data, but rather due to the the presence of a large time interval between data collections. Additionally, this time interval (i.e., night time) is, for many participants, categorically different than the day time. 

There are three ways that state-space modeling can accommodate night effects. First, researchers can use a continuous time state-space approach, which accounts for the large interval between data collections. However, there is an important assumption contained in the use of continuous time models for accounting for night effects: that the state process returns to its stationary distribution over the course of the night. What this implies is a weakening of the temporal dependence between the last observation of a day and the first observation of the next day. Second, an equivalent approach that researchers can use with discrete time state-space models is to include missing observations during the night time period, specifically a number of missing observations that approximate the entire length of the night time period (if there are 5 pings in 12 hours of daytime, there would be 5 missing observations for the night period). What the above approaches do not allow for are specific night effects, such as mood changing after a period of sleep. To account for specific night effects, a third approach is to code time relative to night as with a time trend in any of the above ways that fit the researcher's needs. However, the specification of the specific night effect exogenous variable needs to be considered with respect to the phenomena. For example, how long does the mood decrease due to sleep last throughout the day? This effects how the researcher defines $\mathbf{u}_{t}$ in that it affects where they set their zero point for the time variable.

Generally, understanding and accounting for time trends and/or night effects requires the scientist to consider the lived experience of the participants and how their daily schedules will impact the behavioral outcome. Furthermore, care must be taken to code time so that estimated effects are interpretable across participants. Unfortunately, failing to account for time trends can lead to considerable bias in estimated parameters \citep{shumway_time_2011}. Thus, it is important to handle time trends regardless of whether the time trend is a nuisance unrelated to the study or is of scientific interest. In addition, these time trends may not happen in isolation and can be considered as separate aspects of $\mathbf{u}_{t}$ (i.e. time trends can be modelled alongside observed disruptions). As most time trends can be modelled by including time in the disturbance term $\mathbf{u}_{t}$, most state-space modeling packages can be used. Continuous time modeling of night effects is of course restricted to continuous time analysis packages, such as \texttt{ctsem} and \texttt{dynr}. For accessible reading on this important topic, we again refer readers to \citet{shiffman_ecological_2008}, and to \citet{dethlefsen_formulating_2006, chow_dynamic_2011}.

\subsection{Missing Data}

Missingness is of high importance in EMA data collection, as the protocol itself has high participant burden which, in turn, causes attrition. Researchers must investigate the mechanism of missingness in order to proceed as particular missingingess mechanisms require differing levels of intervention on behalf of the researcher. 

\subsubsection{Missing Data Mechanisms}

Classically, missingness mechanisms have been divided into Missing Completely At Random (MCAR: missingness does not depend on any variable in the model), Missing At Random (MAR: missingness on a variable, Y, depends on another variable in the model, X), and Missing Not At Random (MNAR: missingness on a variable, Y, depends on the observed values of Y). For example, if a researcher is a studying alcohol use, lapses in reporting alcohol use may be caused by another variable under study like employment status (MAR) or, in the worst case, it may be caused by the alcohol use itself (MNAR). This is further complicated by two issues that arise from EMA data. First, the data is temporally related. Second and relatedly, missingness in a timeseries can relate to time, which is what occurs in the cases of time-dependent MAR (TMAR: missingness depends on the time variable) and autoregressive time-dependent MAR (ATMAR: missingness of the outcome is related to previous values of the outcome) \citep{slipetz_missing_2022}. In the former case, it is common to see periodicity in missingness and multiple consecutive points missing (e.g., in a study of drinking, we may see consecutive points missing at night multiple days in a row). There are a number of methods for testing different missingness mechanisms \citep{little_1988, chen_pseudo_2020, jaeger_testing_2006}, though less work has been done on the time dependent mechanisms (TMAR and ATMAR).

In most cases, particularly when the data is MAR, ATMAR, or MNAR, missingness will not be ignorable in EMA data (because of high bias and variability) and will need to be explicitly accounted for. When the data is MCAR or TMAR there is lesser bias and variability in parameter estimates, and thus, it is not as problematic. Hence why particular care must be taken in determining the mechanism of missingness. If missing data is not properly handled, problems with statistical power, validity, and accuracy of parameter estimates will arise. Given the prevalence of missing data in EMA studies and the issues that arise because of it (particularly when there is MAR, ATMAR, or MNAR missingness), it is wise for researchers to have a missing data plan before analysis. 

\subsubsection{Missing Data Solutions}

Missingness can be minimized by study design via compensation schedules, contacting participants, and not overburdening participants with long surveys, though there are a number of approaches to accounting for missing data. Single and multiple imputation approaches have been used, with the \texttt{MICE} \citep{van_buuren_mice_2011} and \texttt{Amelia} \citep{honaker_amelia_2011} $R$ packages being the most developed systems for multiple imputation. Single imputation is when the missing values of a single dataset are imputed with estimated values, while multiple imputation is when multiple versions of the original dataset are imputed with estimated values, each dataset is analyzed separately, and results are pooled across datasets. In recent work, the authors of this manuscript have found that multiple imputation approaches for missing item-level data do not fix bias issues in discrete time state-space models, and that simply using the standard filtering approach (such as the Kalman filter for discrete time state-space models) to impute the state values when observations are missing performs best across a wide variety of conditions \citep{slipetz_missing_2022}. Note that the classical Kalman filter is only appropriate for discrete data, however there are a number of filtering methods that perform the same imputation of state values for continuous time and non-linear state space models. 

Filtering methods are a set of state estimation techniques which sequentially estimate the states and their covariances from the data up to a given timepoint and from the parameters of the model in equations 1-4. As previously mentioned, the classical Kalman filter is appropriate for discrete time state-space with linear dynamics, but any filtering technique can be considered a single imputation method for state values. While there is no hard and fast rule for the maximum of missingness in order for filtering to perform adequately as a missing data correction tool, the authors found that the classical Kalman filter allowed for successfully recovered parameters at 30\% missingness \citep{slipetz_missing_2022}. For missing data in discrete time models, we recommmend using the \texttt{dlm} or \texttt{dynr} packages, as they both natively account for missing data using the Kalman filter. Generally, methods to account for missing data in EMA study analyses is a large and much needed area of study, and we refer interested readers to our work and the work of \citet{padilla_space-time_2020, pratama_review_2016, walani_multiple_2015} for initial investigations of this.

\begin{figure}[H]
\centering
\includegraphics[width=.9\textwidth]{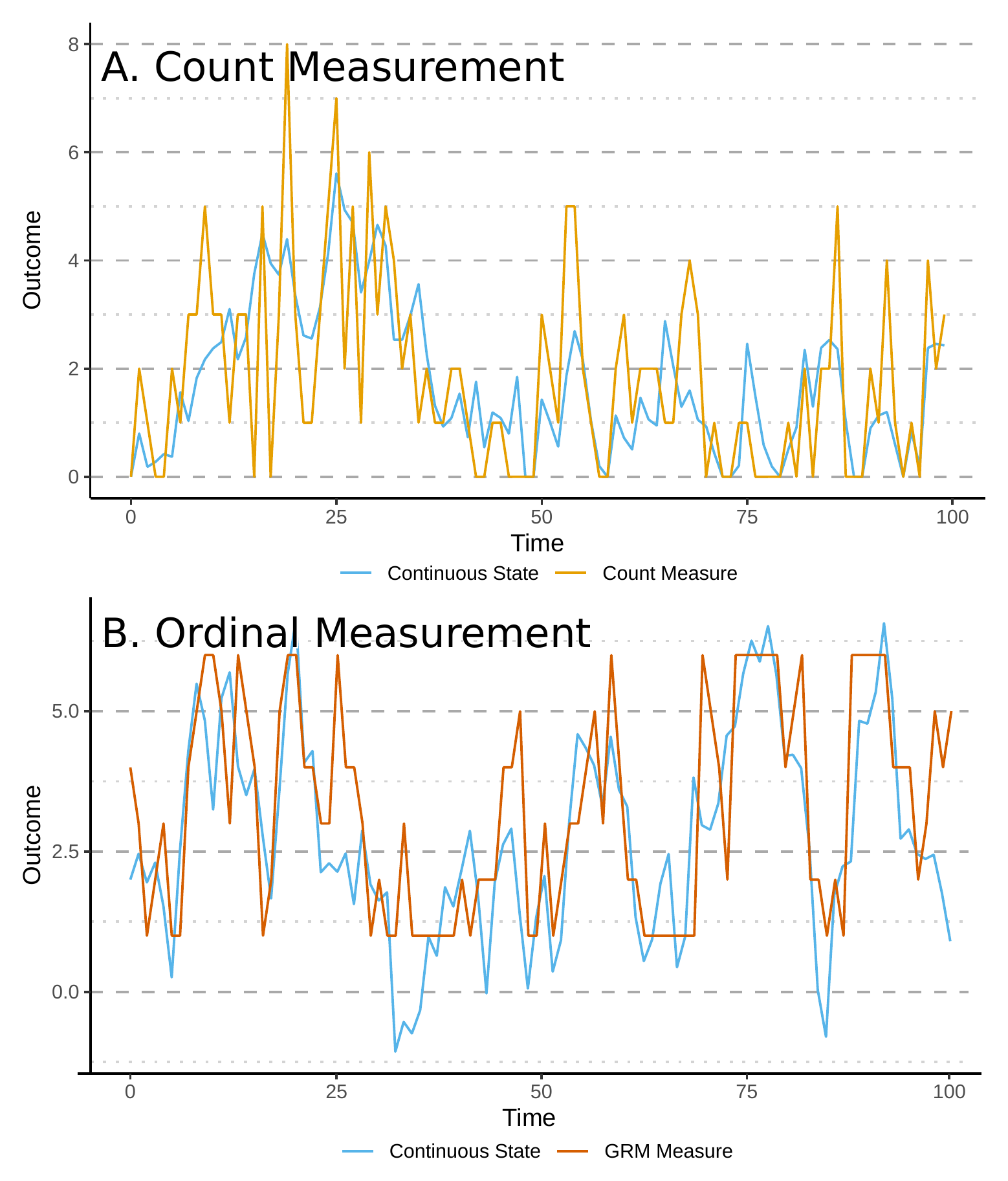}
\caption{Panel A illustrates the underlying state trajectory (blue line) and the corresponding count variable (yellow line) from a count data state-space model, where the count data follows a Poisson distribution. For example, the state trajectory could be the expected count of alcohol use for a single state-space model. Panel B shows a state-space with ordinal measured variables, modeled by an observed graded response model (red line) and the underlying state trajectory (blue line). This model is appropriate for Likert-style measured variables.}
\label{fig2}
\end{figure}

\subsection{Measurement}

State-space models include a measurement model, which maps the unobserved states to the measured observations. The model described in Eqs 1-4 implies that both the states and observations are continuous and normally distributed, which can be best understood as somewhat equivalent to the assumption of the underlying multivariate normality of the latent variables and manifest indicators in latent variable models \citep{bollen_structural_1989}. However, EMA data is rarely collected using truly continuous observations. Instead, observed variables in an EMA study are often Likert scale (e.g., 1-5, agree to disagree), event indicators (0/1 something happened), or some more categorical measurement. Here, taking the mean would not make sense as the distance between each tick (e.g., from 1-to-2 and 2-to-3) is unknown, so the mean would be uninterpretable. From a study design standpoint, researchers have to balance the participant burden against the length of the survey or type of questions they ask at any given ping. One potential way of improving measurement granularity for items that are commonly on a Likert scale is to use sliders, which allow participants to respond on a semi-continuous scale between 0 and 100. Unfortunately, evidence for the use of sliders instead of Likert scale responses is mixed. There is some evidence to suggest that sliders improve measurement \citep{voutilainen_how_2016} (in that there is less of an effect of confounders like age, and there is less of a ceiling effect), but there is other evidence to suggest that sliders don't necessarily improve measurement (in that the appearance of the slider can bias the results) and possibly increase the response time \citep{matejka_effect_2016}, thus increasing participant burden and the aforementioned issue with attrition.

\subsubsection{Count Data}

While converting to sliders is an option for Likert scale items, for many item types used in EMA studies the response categories are not continuously distributed, and cannot, even in theory, be made continuous. Take for example an item about alcohol use. In most cross-sectional research, alcohol use might be measured using a ordinal measure that maps onto binned counts. For example, a common alcohol use item might read: "In the past two weeks, how many drinks on average have you had per day", with the response options being 0: No drinks, 1: 1-3 drinks, 2: 4-5 drinks, and 3: 5+ drinks. These kinds of items are prima facie ordinal, as each response category is greater than the last, while the differences between the response categories are by definition unequal. The use of these kinds of quantity/frequency items to measure substance use have been extensively criticized on the basis of construct validity \citep{mcginley_validity_2014}. For researchers using ordinal binned counts to measure events like substance use, one alternative is to take advantage of the real-time nature of EMA data collection and ask participants to recall their daily use. Instead of resulting in ordinal data, this results in count data (i.e., the number of drinks had during that day). While this is a better practice measurement-wise than using ordinal binned counts, as it avoids combining ranges of counts, the use of count data still requires a change in the measurement component of the state-space model. 

Modifying the measurement component of the state-space model is, in theory, a matter of choosing the correct relation between the unobserved state variables and the observed outcomes. For example, if one considered a count of alcohol use to be distributed according to a Poisson distribution, the resulting (single) state-space model would be
\begin{align}
    \mathbf{x}_{t+1} &= \mathbf{A}\mathbf{x}_t + \mathbf{G}\mathbf{u}_t + \boldsymbol{\varepsilon}_t \\
    \boldsymbol{\varepsilon}_t & \sim N(0, \boldsymbol{\Sigma})\\
    \mathbf{y}_t &\sim \text{Pois}(x_t)      
\end{align}

Here, the state $x_t$ is operationally defined as the expected count of alcohol use per day (as there is only one observed outcome). The underlying state trajectory and corresponding count variable for a system such as this are visualized in Fig. \ref{fig2} Panel A. For researchers that have more than one state variable or count measured variable, we recommend carefully operationalizing the relation between states and measurements, that is, using the same state estimate $x_t$ to represent the average count of alcohol use and cigarette use would ignore the fact that daily cigarette use is typically much higher than daily alcohol use. If a researcher has hypothesized one underlying latent state that drives both alcohol use and cigarette use, at minimum the researcher needs to parameterize the state space model so that the single state value is scaled correctly.  For example, in a single state model with state $x_t$, having both alcohol use and cigarette use represented by a $Poisson(x_t)$ would be incorrect, but allowing for a scaling parameter (i.e. Alcohol Use $\sim Poisson(\beta_1 x_t)$ and Cigarette Use $ \sim Poisson(\beta_2 x_t))$ allows for the proper scaling to be computed. Finally, while we do not explicitly discuss the use of zero-inflated distributions, it is possible to integrate the zero-inflation portion of the distribution into the state-space model.

\subsubsection{Ordinal, Dichotomous and Categorical Measurements}

For naturally ordinal measured variables, such as Likert scales, it is still possible to specify an appropriate measurement model. One measurement model that we recommend for Likert scale variables is the graded response model \citep{samejima_estimation_1969}, which models the cumulative probability of responding in category $k$ or higher as a function of an underlying latent state (in practice, graded response models are used in educational testing, so this underlying state is often referred to as ability). Using our state-space modeling notation for measured variables and states, the measurement model for a single item depending on a single state can be expressed as:

\begin{equation}
    P(y_t > i) = \frac{1}{1+\exp(-\alpha(x_t-\beta_i))}
\end{equation}

where $\alpha$ is the discrimination parameter, controlling how sharply the categories are delineated, while $\beta_i$ is the threshold parameter for the $i$th response category, which denotes the point at which the probability of responding greater than the $ith$ response category is $.5$. For a visualization of the observed graded response model process and the underlying state, see Fig 2, Panel B. The graded response model is appropriate for Likert type responses as it implies that higher latent state values correspond with higher response categories. 

In a similar spirit to the specification of the ordinal measurements, dichotomous measures can be integrated into a state space by using the appropriate measurement model \citep[i.e., a 2 parameter logistic model, an item response theory model where the probability of correctly answering follows a logistic distribution:][]{manapat_2022}. Truly categorical variables are more difficult to integrate into a state-space model as measurements but can be easily integrated as an exogenous predictor of the state via the $\mathbf{u}$ matrix.  

\subsubsection{Estimation}

While it is entirely possible to mathematically propose non-normal measurement models for ordinal, count, dichotomous, and categorical data in the state-space modeling framework, estimating these models are considerably more difficult. The R packages we have listed use Kalman filtering, \texttt{dse} \citep{gilbert_brief_2006} and \texttt{FKF} \citep{luethi_2022}, or one of its many extensions, which are filtering methods that, in the simplest case of traditional Kalman, rely on the assumption of underlying normality of both the states and measured variables. While extended and unscented Kalman filtering, two extensions of Kalman filtering, \citep{simon_j_julier_new_1997, e_a_wan_unscented_2000} can be used to model non-linear dynamics and state-measurement relations, no form of Kalman filtering is appropriate for non-continuous measurement variables.  The few uses of state-space modeling with non-continuous measurement variables either treat ordinal outcomes as continuous \citep{van_rijn_logistic_2005}  or use Bayesian estimation methods \citep{deyoreo_bayesian_2017, imai_bayesian_2009, de_haan-rietdijk_use_2017}, which require customized modeling code to specify the state-space model. From our review of the available tools, there are no ready-to-use software packages for fitting state-space models to non-continuous measurements, with one exception: as of time of writing, the authors of the current manuscript have developed an R package for the simple implementation of graded response model measurement state-space models in a frequentist framework, for use with Likert-type response timeseries commonly seen in EMA data. This package, \texttt{genss}, is publicly available \citep{ami_falk_2023_7887019} and will be extended with other measurement models in the future. 

\subsection{Disturbance Vs Error}
EMA data collection is naturalistic and temporal in nature, collecting data from an individual's real world experience over a given interval. The individual will undergo any number of external events that would impact the phenomenon under study, and while we can consider these events to have a randomly distributed effect, represented in the white noise error term contained in many of the models that are used, there are, of course, events in the lives of participants which will have systematic effects on the process under study. For example, if studying affective processes, day-to-day perturbations in affect likely have a net zero effect, but major life events (bereavement, for example) have systematic effects that, unless modeled, will bias the final models.

Consider the specific meaning of each ``error'' term in Equations 1-4. The innovation term for the state transitions, $\varepsilon_t$, are fluctuations in the underlying state process. This term contains, in the setting of EMA, the collective effect of life events on the state variables, with the strong assumption that the net effect of life events is approximately 0. The error term for the measurement equation $\nu_t$ is measurement error, the degree to which the measurement device is inaccurate, with again the strong assumption that this measurement error averages out to no effect. The first assumption is most problematic in the context of EMA data collection, as it assumes that no life events can cause permanent changes in state processes, only changes that eventually dissipate. 

We label events that cause lasting or major change in the states (and therefore in the measurements) as major disturbances. There are a number of techniques from state-space modeling applied to engineering problems for the detection and removal of disturbances or anomalies, however, the vast majority are difficult if not impossible to apply to EMA data. This is due to the need for these methods to have accurate estimates of both system dynamics and expected error in measurement and state processes, and for disturbances to be true outliers with respect to those models. In EMA data collection and analyses, we have the twin problems that the correct model is unknown, thus needing to estimate it, and that there are an infinite number of major disturbances that can systematically impact the lives of the participants. Thus, the solution to major disturbances in the modeling of EMA data is not to attempt to automatically detect and remove them, but rather to take advantage of the collection modality and collect fine grained information about daily life events. 

\subsubsection{Detecting and Modeling Major Disturbances}

Our recommended solution is the addition of a single daily question that is applicable to most if not all EMA studies: "Did anything memorable, positive or negative, happen today?" If the participant endorses this item, then the protocol should require a brief description of what happened. This question can be tailored to specific outcomes and collection time frames, and simply provides observed data of possible major disturbances. An example of this can be seen in \citet{mackerron_happiness_2013}. While they were interested in the link between well-being and environment, they also collected information on who the participants were with and what they were doing. The key here is to collect information about disturbances via free response, which can then be quantitatively coded into categories for explicit inclusion in the state dynamics equation (Eq 1), specifically in the $\mathbf{u}_t$ disturbance term. However, care must be taken when specifying \textit{how} a disturbance impacts the state dynamics. 

For example, simply coding a life event as impacting a single timepoint (i.e. $\mathbf{u}_t = 1$ for a single $t$) would not result in any sort of lasting effect of that life event, while coding a major life event (i.e., bereavement in a study of depression) as $\mathbf{u}_t = 1 $ for all $t$ after the event occurs implies a lasting effect of that disturbance. A middle ground between these two options would be code an event as a decreasing series of values (i.e., $\mathbf{u}_t = {1, .5, .25, .125, \dots}$ for $t$ after the event occurred). The choice of how to include disturbances in these models must first be informed by the researcher's theoretical understanding of how a disturbance might impact the outcome of interest. We recommend taking an empirical approach when evaluating between several coding options, considering model goodness of fit using an information criteria such as the AIC or BIC under different specifications of the disturbances. It is important to remember that, though not all models will be used, researchers should report the results from all analyses so that readers can assess the sensitivity of the results to the model specification. Multiverse analysis \citep{steegen2016} is also applicable here. 

Fortunately from the implementation side of these models, inclusion of disturbance terms is a simple matter. The previously referenced packages for fitting simple and more complex state-space models, \texttt{dse}, \texttt{ctsem} and \texttt{dynr} all can include disturbances and other exogenous predictors. For more details on the specification of different types of disturbances and interventions, we direct readers to \citet{ryan_time_2022} and \citet{fruhwirth-schnatter_stochastic_2010}.

\begin{figure}[H]
\centering
\includegraphics[width=.9\textwidth]{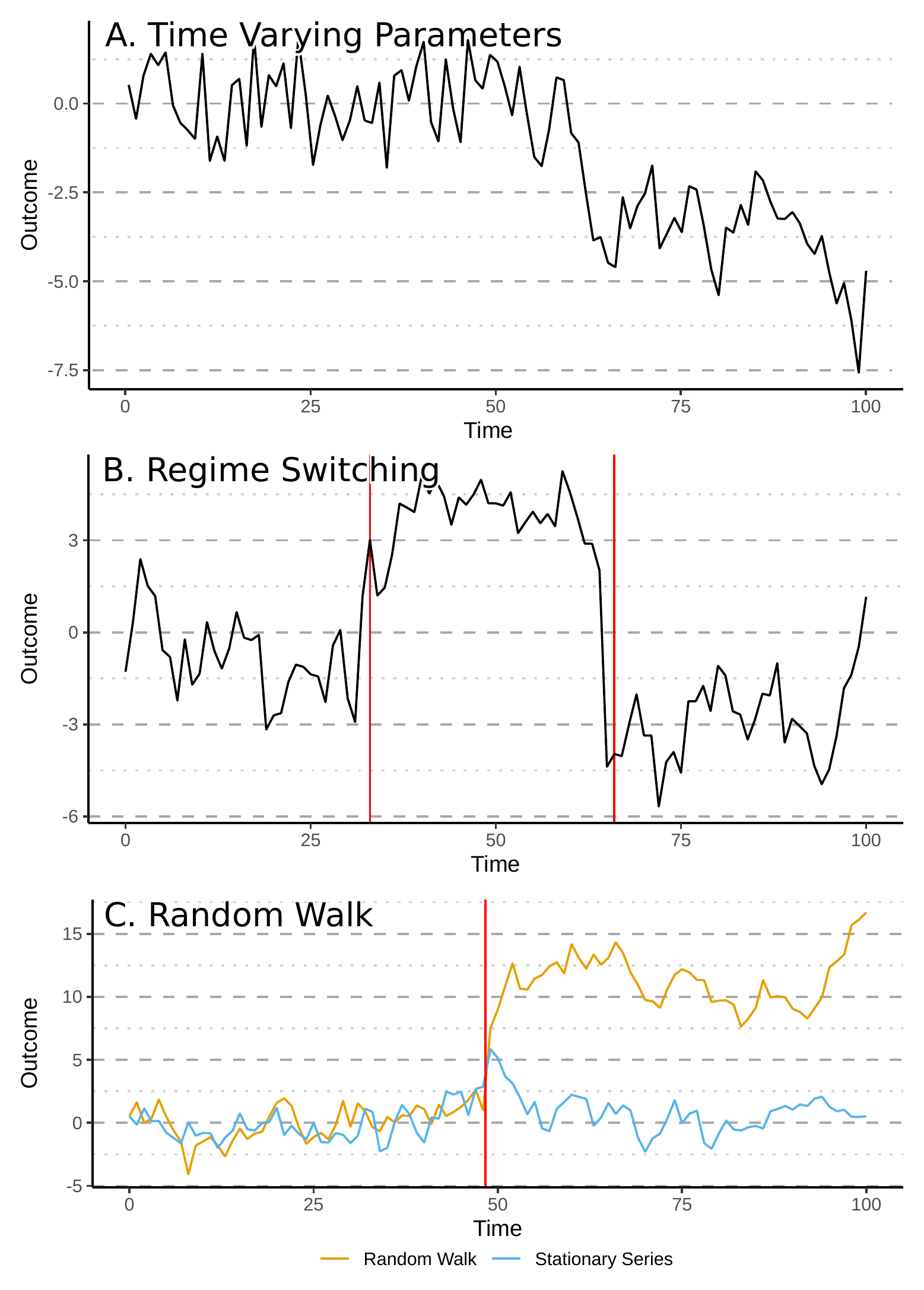}
\caption{This figure illustrates three kinds of non-stationary processes. Panel A is an example of an autoregressive process with a time-varying parameter (the autoregressive parameter starts at 0 for t = 0 and increases in a sigmoidal fashion to 1 at t = 100). Panel B is an example of a regime switching process (the mean changes from 0 to 3 at t = 33 and again from 3 to -3 at t = 66). Panel C is an example of a univariate random walk process (yellow line) and a stationary process (blue line), both with an error variance of $\varepsilon$ and a disturbance at t = 50. For the stationary process, the disturbance causes an increase that returns to fluctuating around the mean of 0. For the random walk process, the disturbance causes an increase and it does not return from the increase.}
\label{fig3}
\end{figure}

\subsection{Stationarity}

The definition of \textit{stationarity} as applied to time series models is complex with a number of technical nuances, but can be broadly summarized as the following: a time series is stationary if the data generating process is the same for every timepoint. While stationarity is a fairly strict property, there are any number of ways a time series might be non-stationary, and we discuss three types of non-stationarity that are most likely to be applicable to psychological data: time-varying parameters, regime switching, and random walks.

\subsubsection{Time-varying parameters}

Time-varying parameters in a state-space framework allow for the parameters governing the state dynamics or measurement properties to vary as a function of time. This captures a number of situations, from the state dynamics $\mathbf{A}$ changing slowly overtime (due to, for example, an intervention being applied) to the meaning of the measurements changing suddenly, resulting in changes to the $\mathbf{H}$ matrix. Figure 3 Panel A shows an example of an autoregressive process with a time-varying parameter. There, the autoregressive parameter starts at 0 when $t = 0$, and increases in a sigmoidal fashion to 1 at $t = 100$. This results in the timepoints that are effectively uncorrelated in time at the beginning of the series, and highly correlated in time at the end of the series. \citet{fisher_square-root_2022} provides an example of a psychological process with potentially time-varying parameters: positive and negative affect before, during and after participants engaged in a mindfulness-based mediation workshop. Using this data, Fisher and colleagues show steady attenuation across the period of the workshop in the autoregressive parameters for both positive and negative affect. This in turn suggests that for some participants, mindfulness based mediation practice could be resulting in the temporal decoupling of affective states, so that both positive and negative affect become less dependent on previous values. Time-varying parameter state-space models can be useful tools for research to explore the impacts of interventions or life events on psychological dynamics; however there are several issues concerning their implementation of which readers must be aware. 

First, as of writing, there are no open-source R implementations of general time-varying parameter state-space models currently available, and previous applications of these models have used bespoke implementations. Second, specifying time-varying parameter models requires careful thought as to which parameters can vary in time and how they vary, as outside of a few special cases, one cannot specify a state-space model where each parameter varies in time arbitrarily. Development and implementation of general time-varying parameter state-space models is an active area of methodological research, and we expect more accessible open source implementations will be developed in the next few years. There is, however, a specific type of time-varying parameter state-space models that has an easily accessible and usable implementation: regime switching state-space models.

\subsubsection{Regime switching}

Regime switching state-space models are a specific type of time-varying state-space model that allow for sudden changes in dynamics or mean levels that remain consistent within a given time interval (also known as a ``regime'') \citep{kim_state-space_1999}. Typically, within a regime the state-space model is stationary, while between regimes the parameters can vary. Regime-switching dynamics can occur in psychological time series for any number of reasons. For example, \citet{chow_dynamic_2011} use regime switching differential equation models to show that the behavior of children in mother-child dyads exhibit sudden shifts in dynamics, from attempts to stay close to the mother to more exploratory behavior. Regime switching is illustrated in Fig \ref{fig3} Panel B. The time series exhibits a change in the mean of the time series at $t = 33$ from 0 to 3, and at $t = 66$ from 3 to -3. Regime switching state-space (or regime switching observed time series models), in either discrete or continuous time are uniquely simple to specify and fit, due to the extensive work of Ou and colleagues in the development of \texttt{dynr} \citep{ou_r_2019}. Researchers should use regime switching models in cases where there are periods of stability, but differences between those periods (e.g., the above case of measuring pre-COVID and during COVID).

\subsubsection{Random walk processes}

Random walk non-stationarity is the one form of non-stationarity discussed here that is not strictly an example of a time-varying parameter model. Put simply, a random walk process is one where changes, be they due to systematic disturbances or regular fluctuations, are permanent. Mathematically, a univariate random walk process can be described as $x_{t+1} = x_t + \varepsilon_t$, while the specification of a multivariate random walk process is more complicated. Figure 3 Panel C illustrates a univariate random walk process (red line) and a stationary process (blue line) with the same error variance of $\varepsilon$. At $t=50$ a disturbance is introduced that raises the value of the series by a fixed constant. Note the differences between the random walk process and the stationary process: The stationary process has an increase at $t = 50$ but then quickly returns to fluctuating around the model implied mean of 0. The random walk process however increases due to the disturbance and stays increased, never returning to the previous level. 

The differences between stationary and random-walk time-series have a number of implications for the modeling of psychological processes. Most significantly, as was previously mentioned in the section on disturbances, the effects of an intervention, treatment or disturbance must be specified to be persistent if the underlying model is stationary, else the effect of the intervention will attenuate and disappear. Miss-specifying the effect of an intervention without considering the impact of stationarity will result in attenuated effect estimates and lower power. Researchers should use random walk processes when the changes that occur in the dynamics of the study are permanent; that is, regime switching would be sufficient for weekend fluctuations in mood, but random walk processes would be best for a major life event, like a death of a loved one. Fitting random walk state-space models can be done by constraining the autoregressive parameter to be, in the case of a single state model, $1$. For multiple states, fitting random- walk models is more difficult, as the auto-regressive and cross-regressive effects must be in balance. For single state models, any state-space modeling package that allows for parameters to be fixed can be used. At time of writing, there is no implementation of multi-state random walk state-space models available.

\begin{table}[]
\centering
\caption{Issues and Recommendations}
\label{tab:recommendations}
\resizebox{\columnwidth}{!}{%
\begin{tabular}{lll}
\hline
\textbf{Issue} &
  \textbf{Recommendation} &
  \textbf{R Packages} \\ \hline
\textit{Idiographic and Nomothetic Modeling} &
   &
   \\ \hline
\begin{tabular}[c]{@{}l@{}}I have no theoretical reasons to believe the phenomenon's\\ dynamics are the same across all individuals.\end{tabular} &
  Use an idiographic (N = 1) approach &
  \begin{tabular}[c]{@{}l@{}}\texttt{dynr}; \texttt{dlm};\\ \texttt{gimme}; \texttt{genss}\end{tabular} \\
\begin{tabular}[c]{@{}l@{}}The   phenomenon is partially idiographic, and my model\\ has latent variables or I am concerned with subgroups.\end{tabular} &
  Use GIMME &
  \texttt{gimme} \\
\begin{tabular}[c]{@{}l@{}}The   phenomenon is partially idiographic, and I have an\\  observed variable model with a large number of parameters.\end{tabular} &
  \begin{tabular}[c]{@{}l@{}}Use a regularized partially idiographic method\\ such as \texttt{multivar}.\end{tabular} &
  \texttt{multivar} \\
\begin{tabular}[c]{@{}l@{}}The   phenomenon is partially idiographic, but I don't have\\ a lot of variables.\end{tabular} &
  \begin{tabular}[c]{@{}l@{}}Use multi-level VAR to allow for \\ inter-individual differences in relations\end{tabular} &
  \texttt{mlVAR} \\ \hline
\textit{Data Collection Schedule} &
   &
   \\ \hline
\begin{tabular}[c]{@{}l@{}}The   phenomenon I'm studying changes on the order of \\ X hours/days/weeks.\end{tabular} &
  Collect responses every X*2 hours/days/weeks &
  Any \\
The   pings are approximately equally spaced. &
  Use a discrete time model &
  \texttt{dynr}; \texttt{dlm}; \texttt{OpenMx} \\
The   intervals between pings are highly varied. &
  Use a continuous time model &
  \texttt{dynr}; \texttt{ctsem}; \texttt{OpenMx} \\
\begin{tabular}[c]{@{}l@{}}I am not   concerned with changes pre-and post-treatment,\\  or I have no mid-study   treatment.\end{tabular} &
  Use a single burst design &
  Any \\
\begin{tabular}[c]{@{}l@{}}I am   concerned about changes in dynamics that occur\\  during the study (due to treatment/interventions)\end{tabular} &
  Use a multiple burst design &
  Any \\ \hline
\textit{Time Trends and Night Effects} &
   &
   \\ \hline
I   believe that there is a time trend in my data. &
  \begin{tabular}[c]{@{}l@{}}Code time as an exogenous variable and include it in\\  $u_t$ (see text for coding options)\end{tabular} &
  Any \\
\begin{tabular}[c]{@{}l@{}}I think that there is a substantial change in the phenomena\\  during the night period (i.e. negative affect decreases during sleep)\end{tabular} &
  \begin{tabular}[c]{@{}l@{}}Code time as an exogenous variable\\ and include it in $u_t$\end{tabular} &
  \texttt{ctsem}; \texttt{dynr} \\ \hline
\textit{Missing Data} &
   &
   \\ \hline
My data   is MAR, MNAR, or ATMAR. &
  \begin{tabular}[c]{@{}l@{}}Characterize and describe missingness issues and\\  use missing data methods to adjust models/data.\end{tabular} &
  Any \\
\begin{tabular}[c]{@{}l@{}}I need a   solution for non-ignorable missingness\\  in a discrete time model.\end{tabular} &
  \begin{tabular}[c]{@{}l@{}}Standard filtering techniques account for\\ missingness well. Compare with established\\ missing data methods like multiple imputation.\end{tabular} &
  \texttt{dynr}; \texttt{dlm} \\ \hline
\textit{Measurement} &
   &
   \\ \hline
I have count data (e.g. substance use count, behavior count) &
  \begin{tabular}[c]{@{}l@{}}Choose the appropriate distribution for \\ the count variable (e.g., Poisson, negative binomial)\end{tabular} &
  Unavailable \\
I have ordinal data (e.g. Likert scale responses). &
  \begin{tabular}[c]{@{}l@{}}Use an ordinal measurement model\\  such as graded response.\end{tabular} &
  \texttt{genss} \\
I have dichotomous data (e.g. symptom presence) &
  Use a logistic/probit measurement model. &
  Unavailable \\ \hline
\textit{Disturbances} &
   &
   \\ \hline
\begin{tabular}[c]{@{}l@{}}There are no specific events/disturbances that this study\\  is explicitly interested in studying.\end{tabular} &
  \begin{tabular}[c]{@{}l@{}}Ask, "Did anything memorable, positive\\  or negative, happen today?"\end{tabular} &
  N/A \\
\begin{tabular}[c]{@{}l@{}}There were disturbances without a lasting effect\\  on the phenomena.\end{tabular} &
  \begin{tabular}[c]{@{}l@{}}Code as single timepoints (u\_t = 1) or short decreasing series\\  (e.g.   u\_t{[}+1,2,3{]} = {[}1,.5,.25{]})\end{tabular} &
  \texttt{dse}; \texttt{ctsem}; \texttt{dynr} \\
\begin{tabular}[c]{@{}l@{}}There is a major disturbance with lasting effects,\\ such as a treatment or intervention.\end{tabular} &
  \begin{tabular}[c]{@{}l@{}}Code as constant for all timepoints\\  after occurrence (e.g. $u_t = 1\text{ for }t >=X$)\end{tabular} &
  \texttt{dse}; \texttt{ctsem}; \texttt{dynr} \\
There is a disturbance whose effect disippates &
  \begin{tabular}[c]{@{}l@{}}Code as a decreasing series \\ (e.g. $u_t[+1,2,3, \dots] = [1,.5,.25, \dots]$)\end{tabular} &
  \texttt{dse}; \texttt{ctsem}; \texttt{dynr} \\
How do I know I'm including the right disturbances? &
  \begin{tabular}[c]{@{}l@{}}Compare models with and without\\ disturbances using AIC or BIC\end{tabular} &
  \texttt{dse}; \texttt{ctsem}; \texttt{dynr} \\ \hline
\textit{Stationarity} &
   &
   \\ \hline
I think the state dynamics are changing over time. &
  Use a time-varying parameter model &
  Unavailable \\
\begin{tabular}[c]{@{}l@{}}I think there are sudden switches in dynamics or means\\ that remain stable over a period of time before switching again.\end{tabular} &
  Use a regime switching model &
  \texttt{dynr} \\
I think there changes to the dynamics or means that are permanent. &
  Use a random walk model &
  Any \\ \hline
\end{tabular}%
}
\end{table}

\section{Summary}

EMA studies have the potential to test and explore a broad variety of time trends in psychological phenomena. In order to do so, one promising framework of analysis is state-space modeling. Using these models, there are a number of considerations that must be taken into account in the design and analysis. Here, we catalogue some of these considerations and offer corresponding recommendations for successful EMA study design and analysis (see Table 1 for a summary of recommendations). These considerations have varying degrees of necessity when planning a study.

\subsection{Study Design}

An EMA study must aim to capture inter-individual differences in the dynamics of the phenomenon under study. Hence, we recommend idiographic or partially idiographic modeling when theoretical reasons preclude a nomothetic model as a nomothetic model that assumes the same process is applicable to all participants in the study can mask the relevant individual dynamics. In addition, time trends must be taken into consideration when modeling the data, with specific care taken to either model the effects of night-time and weekends, or to show that there are no time trends associated with the variables under study. Researchers must ensure that their ping schedule adequately tracks the dynamics of the phenomenon (i.e., a process that changes rapidly throughout the day requires more measurements than a process that remains relatively stable within the day and changes over weeks), while balancing participant burden. Here we suggest that researchers err on the side of improved measurement (i.e. via more items per construct) and integrate practices that improve participant response rate, rather than use short surveys that might have better a priori response rates. We recognize that optimal EMA study design with respect to balancing measurement and burden is highly topic specific, and we encourage research into optimal study design to clarify best practices. Additionally, EMA researchers should have a plan to account for missing data. Recent work suggests that state-space models are fairly robust to missing data due to the ability to use previous timepoints to inform estimates of state at a timepoint with missing indicators \citep{slipetz_missing_2022} as in the Kalman filter; however, this is an unsettled area of methodological research, and we encourage researchers to include a description of the pattern of missingness and an explicit description of any method used to account for missingness in their analyses. 

\subsection{Model Specification}

With regard to state-space model specification, researchers should consider using continuous time models, particularly for designs that measure rapidly fluctuating constructs such as affect over the course of the day. Previous comparisons of discrete time vs. continuous time models for unequally spaced timepoints suggest that discrete time models are robust to a small amount of variability in the time intervals \citep{loossens_comparison_2021}, and continuous time models can be discretized, making them comparable to discrete time models. Even so, care must be taken when interpreting continuous time model's parameters, as effects that have their units in instantaneous rate of change are not directly comparable to effects computed using discrete time models \citep{oud_continuous_2000}. Specifying the correct measurement model is also a concern for EMA researchers, however, the lack of flexible tools for implementing different measurement structures restricts the use of correctly specified models. Until a general solution for fitting non-continuous measurement models in state space models is developed, EMA researchers should consider using quasi-continuous versions of items (i.e. 0-100 sliding scales) instead of ordinal scales with few response options. We do make this suggestion tentatively, as more methodological work needs to be done evaluating how well sliding scales can capture individual differences in the responses (i.e. how granular is too granular for a response scale). Researchers with non-continuous outcome data should note the lack of correct measurement model as a limitation. Finally, EMA researchers should consider the use of non-stationary models, but only when there is a strong theoretical reason to do so. The reason for this caution is that non-stationary models, simply due to their increased complexity, are more likely to overfit the data and require careful interpretation.

\subsection{Limitations and Future Directions}

Due to both the sake of brevity and, also, technical limitations, some considerations and recommendations were excluded from this article. For example, we focused on quantitative (e.g., ordinal measures) instead of more qualitative measures (e.g., open-ended questions) that need to be coded. Interested readers can turn to \citet{mayring_mixing_2007} and \citet{Britt1997ACI}. We also assumed that the underlying state process under study has linear dynamics. It is entirely possible to propose state space models where the state dynamics are non-linear, where the state-error distributions are not normal, and even where the states themselves are non-continuous. For a description of non-linear state-space models see \citet{priestley_non-linear_1988}. The modeling of non-continuous measurement variables in a frequentist framework is on the cutting edge of research. While the authors have begun the first steps with \texttt{genss} package, more development needs to be done. We have identified the following topics as immediate future directions:
\begin{itemize}
    \item \textbf{Non-Normal Measurement Models} - Current accessible implementations of state-space modeling cannot integrate non-normal measurement models. This is a pressing need for psychological researchers, as measurement model misspecification can have a drastic impact on inference.
    \item \textbf{Data Needs for State-Space Models in Psychological/Behavioral Sciences} - As most work on state-space models has occurred in an engineering framework, there has been little methodological work done evaluating the performance of these models under typical situations in psychological science. Studies of how effect sizes, number of timepoints, and number of participants impact statistical power would be of great assistance during study design.
    \item \textbf{Multi-Participant State-Space Models} - While some work has been done on multi-participant timeseries analysis with observed variables, there is little work on how to apply the state-space modeling framework to multiple participants' worth of data. Advances in multi-participant timeseries would be useful as not all studies are purely idiographic.
    \item \textbf{Complex Systems Models for Psychological/Behavioral Data} - One emerging line of thought in psychology is that psychological systems are complex systems. However, our standard models for multivariate timeseries (i.e. ones with linear dynamics) are not capable of capturing any complex system behaviors. Non-linear models allow for complex systems behavior, but require careful specification and construction. More work on the precise nature of "psychological systems are complex systems," both from a psychological theory and quantitative development perspective, is needed. This future direction contains research on non-stationary/non-linear state space models, as well as a call for more theoretical work mapping psychological theory to specific data analysis models. 
\end{itemize}

\subsection{Conclusion}

The state-space modeling framework provides a fruitful avenue for the modeling of timeseries data, particularly EMA. The diversity of topics covered here illustrates the flexibility of the method. With such multiplicious researcher decision points comes the need for researchers to be deliberate and conscientious in their study design and modeling decisions. Overall, our aim for this article was to provide support to EMA researchers in making these choices.  

\bibliography{MLMSEM.bib}
\bibliographystyle{apacite}
\end{document}